\documentclass{ws-procs975x65}

\begin{document}
\title{GRAVITATIONAL FIELDS WITH SOURCES, REGULAR BLACK HOLES,
QUASIBLACK HOLES, AND ANALOGUE BLACK HOLES\footnote{Report of the
Parallel Session AT3 at the Marcel Grossmann Meeting 13, Stockholm 2012,
Proceedings of the Conference.}}

\author{JOS\'E~P.~S.~LEMOS}

\address{CENTRA, Departamento de F\'{\i}sica, 
Instituto Superior T\'ecnico,\\ Universidade de Lisboa - UL,
Avenida~Rovisco Pais 1, 1049 Lisboa, Portugal
\\\vspace{0.1cm}
joselemos@ist.utl.pt}

\author{PAOLO PANI}

\address{CENTRA, Departamento de F\'{\i}sica, 
Instituto Superior T\'ecnico,\\ Universidade de Lisboa - UL,
Avenida~Rovisco Pais 1, 1049 Lisboa, Portugal\\
\&\\
Institute for Theory $\&$ Computation, Harvard-Smithsonian
CfA, 60 Garden Street, Cambridge, MA, USA
\\\vspace{0.1cm}
paolo.pani@ist.utl.pt}

\begin{abstract}

We discuss recent developments in gravitational fields with sources,
regular black holes, quasiblack holes, and analogue black holes,
related to the talks presented at the corresponding Parallel Session
AT3 of the 13th Marcel Grossmann Meeting.

\end{abstract}


%\ccode{PACS numbers: 04.60.Pp}

\bodymatter

\vskip1cm

The AT3 Session of the 13th Marcel Grossmann Meeting covered a variety
of topics which are related to the search of viable gravitational
solutions in the presence of sources containing normal or exotic
matter, and to solutions that take into account corrections to general
relativity.  Most of the talks were devoted to the analysis of black
holes (BHs) in various scenarios. Given the interdisciplinary nature
of the subject some overlap exists with other sessions of the meeting,
in particular with AT2 (Extended Theories of Gravity), AT4 (Modified
Gravity), BH3 (Black Holes) and GT4 (Exact Solutions (Physical
Aspects)).

In view of the multitude of covered topics, we find it useful to
divide this report in separate sections, which also reflect an
internal schedule that was adopted during the meeting.

%%%%%%%%%%%%%%%
\section{Regular black holes, quasiblack holes and wormholes}
%%%%%%%%%%%%%%%
A central topic of the session was related to BH-like solutions of
general relativity and other theories in the 
presence of (standard and
nonstandard) matter fields, such as regular BHs, quasiBHs and
wormholes.

Reinhard Meinel gave an overview of quasiBHs solutions in
Einstein-Maxwell gravity~\cite{Meinel:2011ur,Meinel:2012wm}. 
In the static and
electrically neutral case, Buchdahl showed that self-gravitating
objects cannot reach the BH limit. However, the situation is different
in the case of spinning metrics or in the presence of electromagnetic
fields. In the latter cases the exterior spacetime can approach that
of a Kerr-Newman BH in the quasiBH limit, whereas the interior is a
regular, nonasymptotically flat spacetime with a Kerr-Newman
near-horizon geometry at spatial
infinity. Meinel presented rotating models
with a perfect-fluid, static models with electric charge and solutions
of rotating discs of charged dust.

Jos\'e P. Sande Lemos reported on a series of studies coauthored by Oleg
Zaslavskii~\cite{Lemos:2007yh,Lemos:2010kw} and Vilson
Zanchin~\cite{Lemos:2010te}. Lemos presented the properties that
distinguish a quasiBH from a classical BH, even though for external
far away observers the two spacetimes are virtually
indistinguishable. In brief, these properties are: the
existence of whole
regions of infinite redshift, degeneracy of the spacetime, existence
of outer and inner regions which are impenetrable and dynamically
(rather than causally) disjoint. Despite these peculiar properties,
the curvature invariants remain perfectly regular everywhere in the
quasiBH limit. Lemos concluded by presenting mass and entropy formulae
for these objects and the distinctive features of their Carter-Penrose
diagrams.

In the context of quasiBHs, Vilson Zanchin explored some extremal
limits of charged, polytropic  perfect-fluid spheres in general
relativity~\cite{Lemos:2013xxx}. He constructed numerical solutions
which describe a Reissner-Nordstr\"om exterior and a charged,
perfect-fluid star in the interior. When these solutions are extremal,
the Buchdahl bound can be made arbitrarily close to the BH limit, $M=
R c^2/G$, $M$ and $R$ being the object's mass and radius, respectively.
However, as Zanchin's results showed, in the case of
polytropic stars this requires an infinite polytropic index,
i.e., $\gamma\to\infty$, where the fluid equation of state is defined
as $P=k\,\rho^\gamma$, where $P$ is the fluid pressure and $\rho$ 
is its energy density.

Another important class of BH mimickers is described by Mazur's and
Mottola's gravastars~\cite{Mazur:2001fv,Mazur:2004fk}. Emil Mottola
gave a review of the properties of this
solution. Gravastars are cold, dark compact objects with an interior
de Sitter dark-energy condensate and an exterior Schwarzschild or Kerr
geometry of arbitrary total mass $M$. The compactness
of these solutions can be arbitrarily close to the BH limit, but they
do not possess event horizons, nor singularities. 
Mottola presented the entropy formula for a gravastar
(which scales as $M^{3/2}$ rather than as $M^2$ as in the BH case) and
discussed the role of trace anomalies in generating the quantum
fluctuations necessary to gravastar formation.

Two contributions were on regular BHs. Stefano Ansoldi discussed a
work coauthored by Lorenzo Sindoni~\cite{Ansoldi:2012qv} in which they
present a general procedure to generate a class of (everywhere
regular) solutions of Einstein equations with arbitrary number of
horizons. The general structure of these solutions can be interpreted
as a nested sequence of anisotropic vacua, with necessarily
nonincreasing energy density. Ansoldi discussed current work in
progress to find a suitable classification scheme for the maximal
extension of these multi-horizon regular BHs and to what extent these
solutions can arise in modified gravity scenarios or from
higher-dimensional compactifications. Another important issue to
understand is the stability of the solutions.  
Nami Uchikata discussed
the stability against radial perturbations~\cite{Uchikata:2012zs} of
the regular BH solutions found by Lemos and
Zanchin~\cite{Lemos:2011dq} and their generalization. These solutions
are composed of a single de Sitter core whose surface boundary is a
timelike charged massive thin-shell, whereas the exterior spacetime is
Reissner-Nordstr\"om, with the thin-shell being inside the Cauchy
horizon and without matter pressure~\cite{Lemos:2011dq}.  Uchikata
showed that nearly extremal regular BHs only exist in a certain range
of the ratio $GM/(L c^2)$ (where $M$ is the BH mass and $L$ is the de
Sitter horizon radius) and that some solutions are unstable below a
certain critical value of this ratio.

Two further contributions were devoted to the study of traversable
wormholes in modified gravities. In the standard approach, general
relativity is assumed and wormhole solutions require the presence of
exotic matter that violates some energy condition. An alternative
approach makes use of some modified gravity theory to find traversable
wormholes without invoking any exotic matter.  In this context,
Francisco Lobo reviewed some recent work by himself and collaborators
on traversable wormholes supported by dark
gravity~\cite{Lobo:2010sb}. As he discussed, in alternative theory
scenarios higher-order curvature corrections can be interpreted as a
gravitational fluid, which supports wormhole geometries. The latter
are peculiarly different from their general relativity counterpart.
Along similar lines, Remo Garattini's work discusses self-sustained
traversable wormholes in noncommutative theories of gravity and in
gravity's rainbow models~\cite{Garattini:2011fs,Garattini:2007fe}. In
a semiclassical approach, Garattini computed the energy density of the
graviton one-loop contribution to the energy in a wormhole background
and considered it as a self-consistent source. A
comparison among various models with respect to effective
traversability was also made.

Finally, Eduardo Guendelman presented explorable horned-particle
geometries~\cite{Guendelman:2012sv}, which are indirectly related to
wormhole solutions. More precisely, horned particle spacetimes have a
tube-like structure and, in order to be fully explorable (i.e., not to
possess surfaces of infinite redshift), they require a shell sitting
at the throat with negative surface energy density. The condition of
finite energy of the system (or of asymptotic flatness) implies that
the charged object sitting at the throat completely expels the flux it
produces towards the other side of the horned particle. The geometry
turns out to have a tube-like structure, i.e., the electric flux is
hidden from an outside observer. One important issue to understand is
the origin of the negative energy density at the throat of the horned
particle which can have quantum-mechanical origin.

%%%%%%%%%%%%%%%%%%%%%%%%%%%%%%%%%%%%%%%%%%%%%%%%%%%%%%%%%%%%%%%
\section{Black holes as probes of strong gravity}
%%%%%%%%%%%%%%%%%%%%%%%%%%%%%%%%%%%%%%%%%%%%%%%%%%%%%%%%%%%%%%%
This part was devoted to the study of BHs in modified theories of
gravity (see also Session AT2 chaired by Salvatore Capozziello and
Session AT4, chaired by Fawad Hassan and Shinji Mukohyama).

\subsection{Scalar-tensor and quadratic gravity theories }

Among the most studied extensions to general relativity are theories
which describe fundamental scalar fields nonminimally coupled to
gravity. These theories are well-studied in
cosmology. Among other possibilities, they can be tested against
Einstein's gravity in the strong-curvature regime, e.g., using current
X-ray binary observations and near-future gravitational-wave
signals. The inspiral of two massive BHs is a perfect testbed, because
the system is relatively clean and it probes regions of strong
curvature. Unfortunately, in one of the most studied 
classes of alternatives to
general relativity, the class of scalar-tensor
theories, it is well-known that BH binaries do
not emit scalar radiation to the first Post Newtonian orders. 
During the session, Kent Yagi, Michael
Horbatsch, and Caio Macedo
presented two complementary approaches to circumvent this
obstacle.

Kent Yagi presented a Post-Newtonian formalism to study the evolution
of a compact binary in quadratic gravity~\cite{Yagi:2012gp}. The
latter is an extension of general relativity in which the
Einstein-Hilbert action is supplemented by all independent curvature
invariants to second order in the curvature and coupled to a single
scalar field. In this class of theories, at variance with
scalar-tensor gravity, scalar radiation is emitted from the binary if
at least one of the objects is a BH. This is used to constrain
specific deviations from Einstein's gravity using current
electromagnetic observations and future gravitational-wave
interferometers. Projected bounds from gravitational-wave observations
may place constraints that are more than six orders of magnitude
stronger than those obtained from current solar system observations.

Michael Horbatsch presented a work coauthored by
Cliff~Burgess~\cite{Horbatsch:2011ye}, in which they show that scalar
dipole radiation can be actually emitted by a BH binary in
scalar-tensor theories if the scalar field is slowly evolving in time,
e.g., due to a cosmological evolution. The analysis is based on
a previous
result by Jacobson~\cite{Jacobson:1999vr}, who showed the existence of
BHs with time-dependent scalar hairs. In this case, the scalar dipole
emission can be constrained using quasar observations, putting bounds
on the cosmological evolution of light scalar fields.

In the context of quadratic gravity theory considered by Yagi,
Macedo presented analytical solutions describing slowly-rotating BHs
in this theory~\cite{Pani:2011gy}. These solutions include, as
particular cases, nonKerr BHs that arise as consistent solutions of
some modified gravity and that are known in closed form. Macedo
presented parametric corrections to the geodesic quantities (e.g., the
frequency of the innermost stable circular orbit) which can be useful
to constrain the theory via electromagnetic observations. Furthermore,
having an analytical expression for the BH metric greatly facilitates
other types of analysis, e.g., computing the linear response of
spinning BHs and the corresponding gravitational-wave emission.

\subsection{Spinning BHs and tests}

The interaction of spinning BHs and matter fields
might give rise to very interesting effects. This is the case of the
so-called BH bombs, first studied by Press and
Teukolsky~\cite{Press:1972zz}. In a further contribution, Paolo Pani
presented a new framework to study perturbations of slowly-rotating
BHs and applied it to the study of vector fields around Kerr
BHs~\cite{Pani:2012vp,Pani:2012bp}. As it turns out, massive spin-1
fields are subject to BH superradiance precisely as scalar fields do,
giving rise to a BH bomb instability. Pani used this result to
constrain the mass of possible spin-1 particles from current
observations of spinning supermassive BHs.

Furthermore, Dinesh Singh described the dynamics of a spinning test
particle in a BH spacetime, taking into account spin-curvature
coupling effects through the Mathisson-Papapetrou-Dixon
equations~\cite{Singh:2008qr}. As Singh discussed, it is in principle
possible to use this framework as a diagnostic tool to probe the
structure of curved spacetime in suitable astrophysical contexts, for
example when the test particle is taken to be a solar mass compact
object orbiting an intermediate-mass BH. As an illustration, Singh
considered the inspiral around a Kerr BH and that around a nonspinning
but dynamical BH during its late-time ringdown phase (the latter
modelled by a Vaidya metric). The gravitational-wave signal emitted by
the system in the two cases carries unique information on the nature
of the central object, providing another means to search for
signatures of BH mergers and to test the no-hair theorem in general
relativity.

\subsection{Singularities in alternative theories}

The contributions were also
devoted to the singularity problem,
namely, how corrections to general relativity can avoid the
singularities that quite generically appear in Einstein's gravity. Two
classes of theories that resolve curvature singularities under certain
circumstances are Palatini $f(R)$ gravity and
the Eddington-inspired Born-Infeld
gravity. A common feature of these theories is
that they are equivalent to general relativity in vacuum, but
introduce nonlinear couplings to the matter fields. The latter are
eventually responsible of resolving the singularities in early-time
cosmology and in the stellar collapse.

T\'erence Delsate presented a work coauthored by Jan
Steinhoff~\cite{Delsate:2012ky}, in which they uncover the mechanism
responsible of the singularity resolution in this class of
theories. They also show that there exists a degeneracy between
corrections due to the nonminimal couplings and the matter equation of
state. Delsate showed that this degeneracy must be disentangled in
order to constrain these theories. Interestingly enough, the very same
mechanism that prevents the appearance of singularities seems to give
rise to other kinds of pathologies~\cite{Pani:2012qd}.

Gonzalo Olmo presented spherically-symmetric, charged BH solutions in
Palatini gravity~\cite{Olmo:2012nx}, a work coauthored by Diego
Rubiera-Garcia. They found that charged BHs in this theory have a
central core whose area is proportional to the Planck area times the
number of charges, whereas far from the core they approach the
standard Reissner-Nordstr\"om metric. However the causal structure of
these solutions is very different from their general relativity
counterpart. Indeed, several interesting solutions exist, including
regular BHs and solutions with a single nondegenerate horizon.

Finally, Vincenzo Vitagliano discussed the cosmological appearance of
dynamical horizons in an inhomogeneous universe in the context of
Brans-Dicke theory, a particular subclass of scalar-tensor
gravity theories~\cite{Faraoni:2012sf}. 
Vitagliano presented a two-parameter
family of spherically symmetric and time-dependent solutions of
Brans-Dicke cosmology, which should describe central objects embedded
in a spatially flat universe. There exist multiple dynamical
apparent horizons, both BH horizons covering a central
singularity and cosmological ones. In some cases, these horizons can
dynamically merge, leaving a naked singularity enclosed in a
cosmological horizon.

%%%%%%%%%%%%%%%%%%%%%%%%%%%%%%%%%%%%%%%%%%%%
\section{Nonasymptotically flat gravitational fields}
%%%%%%%%%%%%%%%%%%%%%%%%%%%%%%%%%%%%%%%%%%%%
While most of the session was devoted to asymptotically flat
spacetimes in the presence of sources,
also solutions with matter 
sources containing a cosmological constant have been discussed.
Two contributions of the session were devoted to black branes,
i.e., black objects whose horizon topology is planar~-- in 
the presence
of a negative cosmological constant, and another
one was devoted to cylindrically
symmetric spacetimes.

Kengo Maeda presented static,
inhomogeneous charged black-brane solutions of Einstein-Maxwell
gravity in asymptotically anti-de Sitter
spacetime~\cite{Maeda:2011pk}. Such solutions are constructed
perturbatively (both numerically and analytically) starting from
Reissner-Nordstr\"om black branes. Interestingly, Maeda shows that the
Cauchy horizon disappears in these solutions, a result that supports
the strong cosmic censorship conjecture. In the extremal limit, even
if the curvature invariants are small close to the horizon, a
free-fall observer would experience infinite tidal forces at the
horizon for any long wavelength perturbation.

Mariano Cadoni presented black-brane solutions of
Einstein-Klein-Gordon theory endowed with a scalar
hair~\cite{Cadoni:2011yj}. The scalar potential is such that these
theories admit anti-de Sitter vacuum. On the other hand, well-known
no-hair theorems forbid the existence of hairy black-brane solutions
with anti-de Sitter asymptotics in this model. By relaxing the
requirement of an asymptotic anti-de Sitter solution, Cadoni showed
that these models allow for hairy black-brane solutions with
non-anti-de Sitter domain wall asymptotics, whose extremal limit is a
scalar soliton. He also discussed several features that make these
solutions particularly interesting for holographic applications, for
example in the context of hyperscaling violation in critical systems.

As another case of solutions with nonflat asymptotics, Irene Brito
presented cylindrically symmetric solutions with a cosmological
constant that match a conformally flat interior to an exterior
Linet-Tian spacetime~\cite{Brito:2012mc}. Brito showed that such
matching is only possible when the cosmological constant is
positive because of the existence of an upper limit on the mass density
of these solutions. In the case of vanishing or negative cosmological 
constant, the mass would exceed such limit.

%%%%%%%%%%%%%%%%%%%%%%%%%%%%%%%%%%%%%%%%%%%%% 
\section{Analogue models
of gravity and analogue black holes}
%%%%%%%%%%%%%%%%%%%%%%%%%%%%%%%%%%%%%%%%%%%%% 

In this 13th edition of
the Marcel Grossmann Meeting, the contributions about analogue BHs have
been merged with the rest of the topics of the AT3 Session.

Among various analogue models, acoustic models have attracted particular
attention. Under certain conditions, the motion of sound waves in a
fluid is governed by a Klein-Gordon equation propagating on an
effective acoustic metric, which is determined by the flow
properties~\cite{Unruh:1980cg,Visser:1997ux}.

In this context, Leandro Oliveira presented a
work~\cite{Dolan:2010zza}, coauthored by Dolan and
Crispino, in which they compute the quasinormal modes of a
draining bathtub. The latter is a circulating, draining flow whose
acoustic metric shares many key features of a Kerr spacetime. The
linear perturbation equations are evolved in the time domain and also
solved in the frequency domain. Oliveira showed that these spacetimes
present the typical quasinormal ringing~\cite{Berti:2004ju} and that
the characteristic modes can be interpreted geometrically in terms of
null geodesic orbits.

A related contribution was presented by Ednilton Oliveira, who
discussed the propagation of plane waves on a draining bathtub
vortex~\cite{Dolan:2011zza}. The dynamics exhibits an analogue of the
Aharonov-Bohm effect in quantum mechanics. As suggested,
this effect can be observed in the laboratory using gravity waves in a
shallow basin. Oliveira presented a modified version of the effect,
which is inherently asymmetric even in the low-frequency limit. This
leads to novel interference patterns which carry the signature of both
rotation and absorption.

Finally, Christian Cherubini presented a recent
study~\cite{Cherubini:2011zz} which uncovers a relation between analog
gravity and the nonlinear von Mises wave equation of fluid
dynamics. Interestingly, such correspondence is valid at any
perturbative order. Cherubini also focused on the canonical draining
bathtub configuration, discussing the results of some 2+1 numerical
simulations. He discussed the acoustic analogue of superradiance and BH
bomb effects in these geometries~\cite{Berti:2004ju}. 
He concluded by extending the
analysis to include compressibility effects.

%%%%%%%%%%%%%%%%%%%%%%%%%%%%%%%%%%%%%%%%%%%%%%
\section{Conclusions}
%%%%%%%%%%%%%%%%%%%%%%%%%%%%%%%%%%%%%%%%%%%%%%
Several interesting issues and 
ideas have been raised during the discussion.

As for regular BHs, quasiBHs and other BH mimickers, an important
issue is related to their stability and formation. Can these objects
be formed in viable astrophysical scenarios? Are they (at least
linearly) stable? To answer the latter question, recently a linear
stability analysis of regular BHs has been
performed~\cite{Flachi:2012nv}. Formation in astrophysical viable
scenarios remains an interesting open question.

As for modified theories of gravity and their astrophysical imprint,
the most relevant question is to what extent such deviations are
observable in the near future and how to select
viable theories from the plethora of modifications of general 
relativity that have
proliferated during the last
decades. Electromagnetic observations of the
galactic center~\cite{Gillessen:2008qv,Johannsen:2010xs} (and
especially gravitational-wave observations of BH binaries) will
certainly be instrumental in the near future and they will provide
complementary information with respect to observations at cosmological
scale and those involving the structure of
compact stars~\cite{Pani:2011xm}. The latter suffer from the problem
that possible deviations from general relativity are degenerate with different
equations of state at nuclear density, whose microphysics is not
completely understood.

In the context of anti-de Sitter solutions --~beside the stability
issue~-- the most urgent question is to understand whether theories
that admit solutions with nontrivial hairs have a solid holographic
interpretation and how to characterize the dual theory. Scalar dressed
solutions are particularly relevant in the context of the so-called
anti-de Sitter/condensed matter correspondence~\cite{Hartnoll:2009sz}.

As for analogue gravity, in recent years real experiments have been
performed to test some of the predictions of analogue
models~\cite{Weinfurtner:2010nu}, and other experiments have been
designed. In the short term, the experimental challenges will be
overcome and the field will witness an exciting interplay between
theoretical predictions and accumulating experimental
evidences~\cite{Weinfurtner:2013cd}.

%%%%%%%%%
\section*{Acknowledgments}
%%%%%%%%%
JPSL thanks FCT-Portugal for financial support 
through the projects PTDC/FIS/098962/2008
and PEst-OE/FIS/UI0099/2011.  
PP thanks support from
the Intra-European Marie Curie contract
aStronGR-2011-298297 and from FCT-Portugal through PTDC projects
FIS/098025/2008, FIS/098032/2008, CERN/FP/123593/2011.

%\begin{verbatim}

%\end{verbatim}

%\begin{verbatim}
%\bibliographystyle{ws-procs95x65}
%\bibliography{ws-pro-sample}
%\end{verbatim}

\end{document}